# Auxiliary Calculations for Graphene-Based Quantum Hall Arrays Using Partially Recursive Star-Mesh Transformations

D. S. Scaletta[1] and A. F. Rigosi[2,a]

[1]*Department of Physics, Mount San Jacinto College, Menifee, California 92584, USA*

[2]*Physical Measurement Laboratory, National Institute of Standards and Technology (NIST), Gaithersburg, Maryland 20899, USA*

[a] Author to whom correspondence should be addressed. email: afr1@nist.gov

A previous mathematical approach adopted for optimizing the number of total device elements required for obtaining high effective quantized resistances in graphene-based quantum Hall array devices (QHARS) has been further explored with partial recursion patterns. Designs would assume the use of epitaxial graphene elements, whose quantized Hall resistance at the $\nu = 2$ plateau ($R_\text{H} \approx 12906.4~\Omega$) becomes the building block for larger effective, quantized resistances. Auxiliary calculations suggest the importance of applying full recursions at least once to maximize the reduction of total QHARS elements needed for high resistances.

Graphene, when grown epitaxially (EG), has been developed into devices for electrical metrology due to its robust quantum Hall effect (QHE) [1-3], and most EG-based devices that are used as resistance standards operate at the resistance plateau formed by the $v = 2$ Landau level ($R_\mathrm{H} = \frac{1}{2}\frac{h}{e^2} \approx 12906.4037\ \Omega$). This common single-value constraint severely limits the infrastructure and equipment with which one may disseminate the unit of the ohm. Two dominant approaches to remove these limitations include: (1) quantum Hall array resistance standards (QHARS) to link multiple Hall elements in parallel or series, and (2) the use of *p-n* junctions, with both approaches yielding resistances of $qR_\mathrm{H}$ where $q$ is a positive rational number [4-17]. Generally, standardization of electrical quantities has a broad importance to physics, including measurements of electrical properties like resistivity and Hall effect, and enable precise understanding of a material's electronic structure and behavior [18-21].

Device design alternatives must be explored since there are limitations on the total number of feasibly attainable QHARS elements, as explained in [22]. For instance, the maximum achievable quantized resistance from having 500 elements in series is approximately 6.5 MΩ, which is much smaller than the range of resistances currently calibrated globally – up to PΩ levels in some cases [23]. Future QHARS devices may use recursive star-mesh transformations that can achieve resistances at the highest levels of necessity [24-28]. As a supporting calculation to Ref. [22], this work further explores the framework for utilizing star-mesh QHARS device designs by using partial recursions for the sake of understanding how such systems behave and to show practical design considerations that allow for more flexibility in accessing custom quantum resistance values.

Here we will recall some of the fundamental principles and conclusions of Ref. [22]. The mathematical relationship between a star network and its equivalent mesh network ($N$ is equal in both networks, but the mesh contains one fewer node) is:

$$R_{ij} = R_i R_j \sum_{\alpha=i}^{N} \frac{1}{R_\alpha}$$

(1)

In Eq. 1, the indices go as high as $N$ with the condition that $i \neq j$. To simplify how a QHARS device undergoes minimal-element design optimization, let us define $q \equiv \frac{R}{R_\mathrm{H}}$, where $q$ is defined as the number of single Hall elements held at the $v = 2$ plateau to obtain the total resistance $R$. Note that this coefficient $q$, the *coefficient of effective resistance* (CER), is restricted to the set of positive integers ($q: q \in \mathbb{Z}^+$).

One may rewrite Eq. (1) as the following expression:

$$q_{ij} = q_i q_j \sum_{\alpha=i}^{N} \frac{1}{q_\alpha}$$

(2)

And with the work from Ref. [22], including all definitions presented therein, the parameter *M*, or recursion number, was applied to the subscripts in such a way that actual number of elements is represented by $q_{M:i}$ (single index) and the effective number of elements is represented by $q_{M:ij}$ (two indices). We recall:

$$q_{M:i} = \frac{1}{\xi}(\xi q_{M:ij} + 1)^{2^{-M}} - \frac{1}{\xi}$$

(3)

And that the total number of elements in the final QHARS device is:

$$D_T(M, \xi, q_{M:ij}) = 2^M q_{M:i} + \sum_{x=1}^{M} 2^{x-1} \xi = \frac{2^M}{\xi}(\xi q_{M:ij} + 1)^{2^{-M}} - \frac{2^M}{\xi} + (2^M - 1)\xi$$

(4)

This function of three variables was used as the starting point for a final optimization process. The difference in the approach, at this point, will be in modifying the definition of *M*. Therefore, we expect Eqs. 3 and 4 to change based on the definition. This work will consider a few different definitions of *M*, starting with the one in Fig. 1. (Recall that the original definition of *M* implied that each branch would be split via star-mesh transformation, whereas this work deals with subsets only).

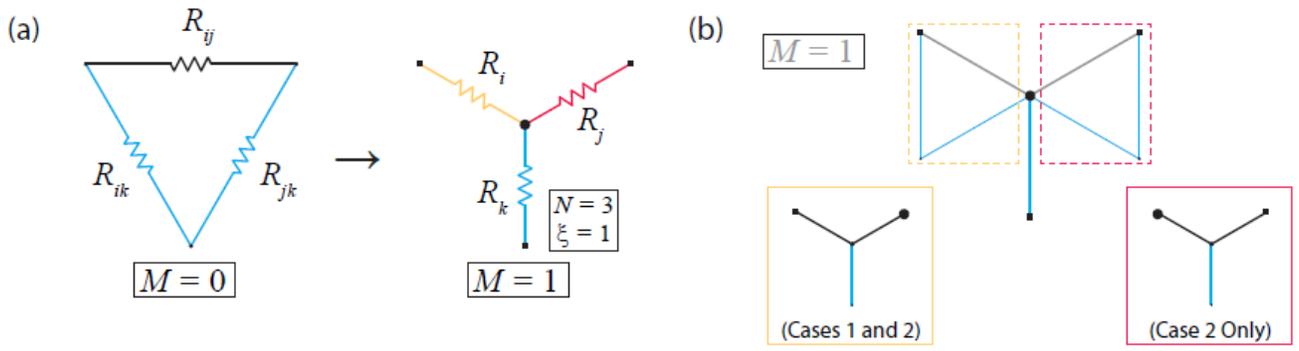
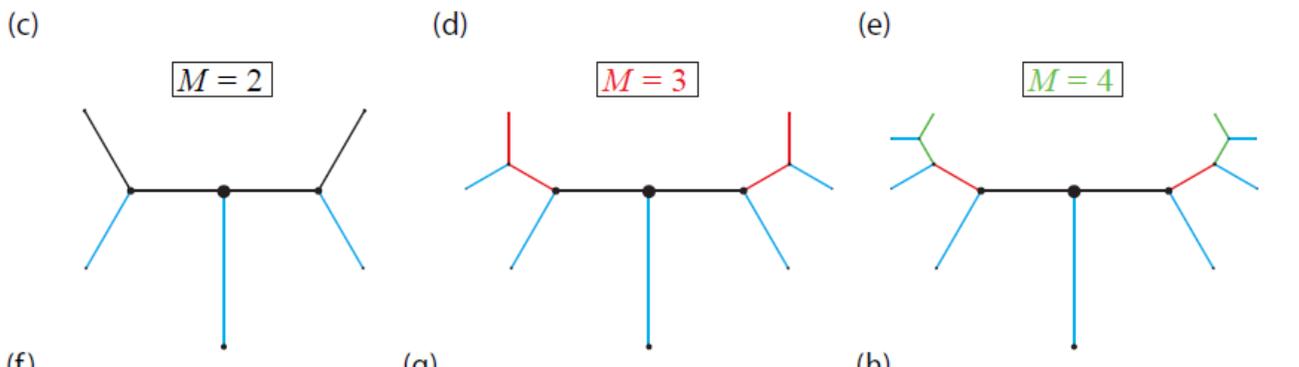
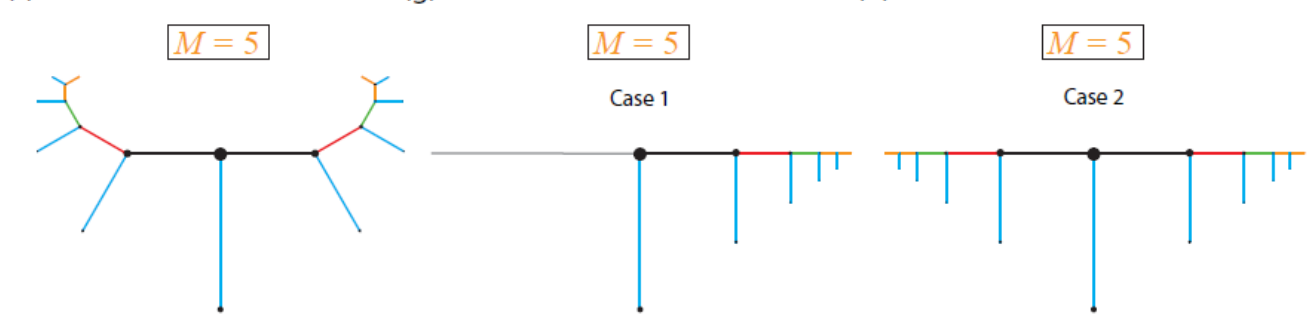
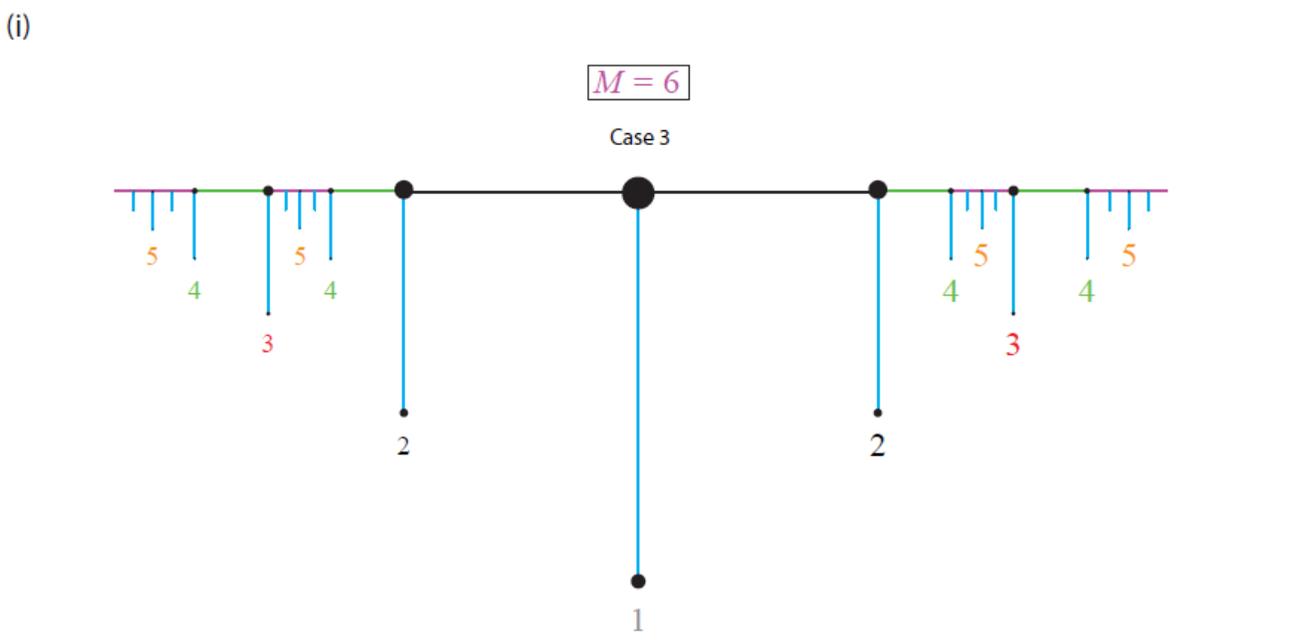

FIG. 1. (a) $R_{ij}$ (equivalently, $q_{ij}$) as a Y-Δ network. Every subsequent expansion of some subset of non-grounded elements increases the characteristic recursion factor $M$ by one. (b) Every resistor in the non-grounded path is expanded as a Y-Δ network, with the full substitution shown in (c). The recursions (for Case 2) are applied for (d) $M = 3$, (e) $M = 4$, (f) $M = 5$, and are done so to grasp the symmetric nature of the recursion. (g) Case 1 ($M = 5$) and (h) Case 2 ($M = 5$) are shown in alternate, topologically similar diagrams used in Ref. [22] for simplicity. (i) Case 3 is drawn up to $M = 6$, where only even numbered recursions are applied fully from the previous (odd) recursion. For odd numbered recursions, Case 2 is replicated.

For the first case, where $M$ indicates the recursive expansion of only one of the two existing branches most previously dealt with, we should introduce the notation $q_{M:i}^{(x)}$, with a superscript $(x)$ to mark the level of recursion we are considering (see Fig. 1). This notation does not change how one calculates the number of elements for a particular branch:

$$q_{M:i}^{(x)} = \frac{1}{\xi}(\xi q_{M:ij} + 1)^{2^{-x}} - \frac{1}{\xi}, x = \{1, 2, \dots, M\}$$

(5)

However, the total number of elements in a QHARS device changes substantially:

$$D_T(M, \xi, q_{M:ij}) = M\xi + \frac{1}{\xi}(\xi q_{M:ij} + 1)^{2^{-M}} - \frac{1}{\xi} + \sum_{x=1}^{M} \frac{1}{\xi}(\xi q_{M:ij} + 1)^{2^{-x}} - \frac{1}{\xi}$$

(6)

And this simplifies to:

$$D_T(M, \xi, q_{M:ij}) = M\xi - \frac{(M+1)}{\xi} + \frac{1}{\xi}(\xi q_{M:ij} + 1)^{2^{-M}} + \frac{1}{\xi}\sum_{x=1}^{M}(\xi q_{M:ij} + 1)^{2^{-x}}$$

(7)

Unlike Eq. 4, which is the main result of Ref. [22], Eq. 7 is discretized as a sum the depends on $M$. To perform similar optimization processes as with the full recursive case, it may benefit the optimizer by temporarily making $D_T$ a smooth function once $q_{M:ij}$ has been selected. The examples in [22] of 1 EΩ, 1 PΩ, 1 TΩ, 1 GΩ will be used for approximating $q_{M:i}^{(M)}$ and $D_T$, with focus on plots that set $\xi$ as 1, 2, 3, 5, and 10.

For the second case, $M$ indicates the recursive expansion of only the outermost existing branches (see Fig. 1). Repeating the analysis yields the following two formulas:

$$q_{M:i}^{(x)} = \frac{1}{\xi}(\xi q_{M:ij} + 1)^{2^{-x}} - \frac{1}{\xi}, x = \{1, 2, \dots, M\}$$

(8)

Due to symmetry, Eq. 8 is identical to Eq. 5.

$$D_T(M, \xi, q_{M:ij}) = (2M - 1)\xi + \frac{2}{\xi}(\xi q_{M:ij} + 1)^{2^{-M}} - \frac{2}{\xi} + 2\sum_{x=2}^{M}\left[\frac{1}{\xi}(\xi q_{M:ij} + 1)^{2^{-x}} - \frac{1}{\xi}\right]$$

(9)

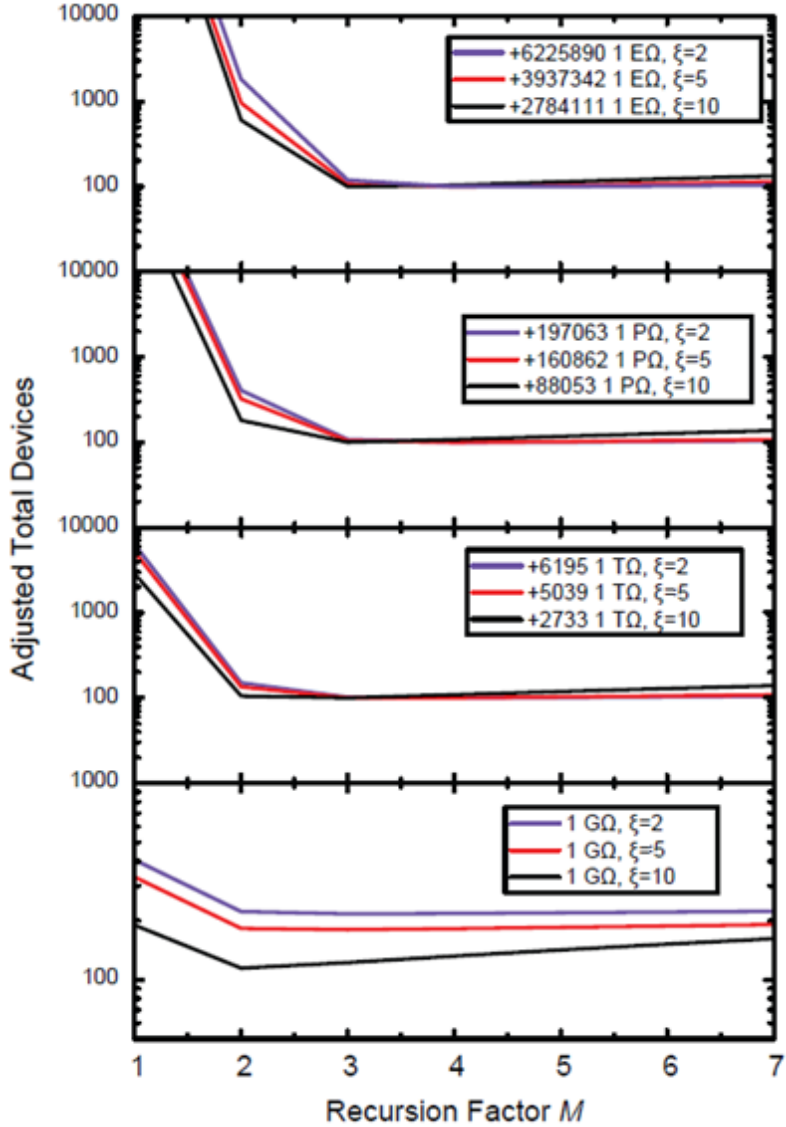

FIG. 2. Calculations for the adjusted total devices needed for specific star-mesh configurations. Resistances from top to bottom include 1 E$\Omega$, 1 P$\Omega$, 1 T$\Omega$, and 1 G$\Omega$. Within each panel, $\xi$ takes on the values 2, 5, and 10. The first three top panels are logarithmically normalized such that the minimum values are at 100 for easier comparison. The true total number of devices is 100 added to the corresponding values in each of the panel legends (except the fourth, bottom, panel that is exact).

For the third case, $M$ indicates an alternation of two partially recursion patterns as seen in Fig. 1, meant to be a hybrid case of a full recursion and the second case. Repeating the analysis yields the following two formulas:

$$q_{M:i}^{(x)} = \frac{1}{\xi}\left(\xi q_{M:ij} + 1\right)^{2^{-x}} - \frac{1}{\xi}, x = \{1, 2, \ldots, M\}$$

(10)

Due to symmetry, Eq. 10 is identical to Eqs. 5 and 8. The total device count becomes unusual due to the alternation:

$$D_T(M, \xi, q_{M:ij}) = 2\xi \sum_{x=1}^{M}\left[2^{\frac{x-H[(-1)^{x+1}]}{2}-1}\right] + \left(\frac{1}{\xi}(\xi q_{M:ij}+1)^{2^{-M}} - \frac{1}{\xi}\right) * 2^{\frac{M+H[(-1)^{M+1}]}{2}+H[(-1)^M]} + H\left[M-\frac{3}{2}\right]$$
$$* \sum_{x=1}^{M-1} 2^{\frac{x}{2}} * H[(-1)^x] * \left(\frac{1}{\xi}(\xi q_{M:ij}+1)^{2^{-x}} - \frac{1}{\xi}\right)$$

(11)

In Eq. 11, $H[x]$ is the Heaviside function. Conventionally, the Heaviside function is used to introduce a binary multiplicative operation on various terms to match the alternating nature of the case. For calculations, the function may be substituted by an approximate analytical version with a large corresponding exponent: $H[x] \approx \frac{1}{1+e^{-2kx}}$

Figure 2 shows the resulting calculations for all three cases for a specific desired resistance. Calculations for the adjusted total devices needed for specific star-mesh configurations. Resistances from top to bottom include 1 E$\Omega$, 1 P$\Omega$, 1 T$\Omega$, and 1 G$\Omega$. Within each panel, $\xi$ takes on the values 2, 5, and 10. The first three top panels are logarithmically normalized such that the minimum values are at 100 for easier comparison. The true total number of devices is 100 added to the corresponding values in each of the panel legends (except the fourth, bottom, panel that is exact).

Figure 3 shows calculations for the adjusted total devices needed for 1 E$\Omega$. From top to bottom, each panel reflects Case 1, 2, and 3, respectively. Within each panel, $\xi$ takes on the values 2, 5, and 10. The logarithmic normalization is applied again for easier comparison. The true total number of devices is 100 added to the corresponding values in each of the panel legends.

What can be surmised from the three cases is that even by including additional branches of potentially infinite recursion, what makes the largest contribution to the total device count is the very first iteration of the resistance network. That is, it always benefits a device designer that $M$ is equal to or greater than 2. More auxiliary calculations can be performed to further understand the impact of partial recursions.

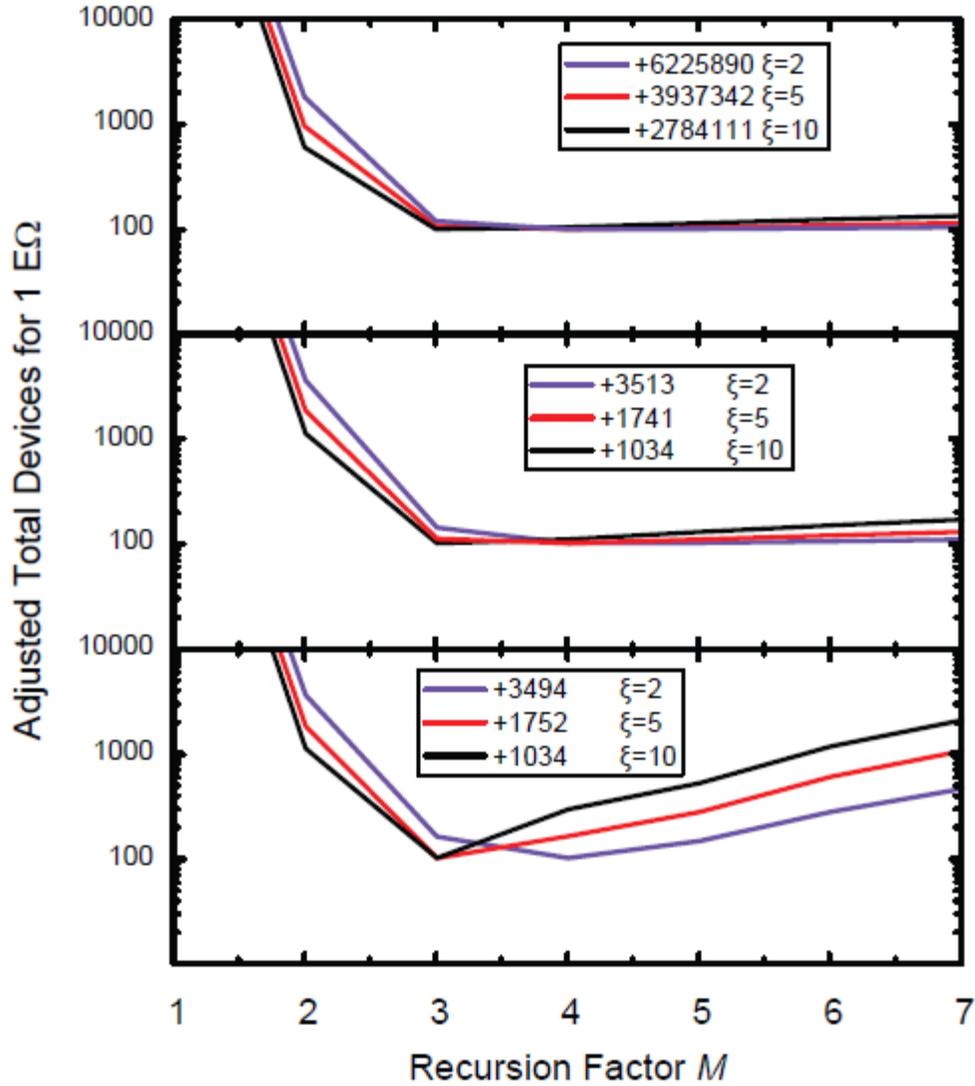

FIG. 3. Calculations for the adjusted total devices needed for 1 EΩ. From top to bottom, each panel reflects Case 1, 2, and 3, respectively. Within each panel, ξ takes on the values 2, 5, and 10. The logarithmic normalization is applied again for easier comparison. The true total number of devices is 100 added to the corresponding values in each of the panel legends.

## ACKNOWLEDGMENTS

The authors thank N. T. M. Tran, F. Fei, G. J. Fitzpatrick, and E. C. Benck for their assistance in the NIST internal review process. The authors declare no competing interests. Commercial equipment, instruments, and materials are identified in this paper in order to specify the experimental procedure adequately. Such identification is not intended to imply recommendation or endorsement by the National Institute of Standards and Technology or the United States government, nor is it intended to imply that the materials or equipment identified are necessarily the best available for the purpose.

# REFERENCES

[1] M. Kruskopf, S. Bauer, Y. Pimsut, A. Chatterjee, D. K. Patel, A. F. Rigosi, R. E. Elmquist, K. Pierz, E. Pesel, M. Götz, J. Schurr, Graphene quantum Hall effect devices for AC and DC electrical metrology. IEEE Trans. Electron Dev. **68**, 3672, (2021).

[2] M. Marzano, M. Kruskopf, A. R. Panna, A. F. Rigosi, D. K. Patel, H. Jin, S. Cular, L. Callegaro, R. E. Elmquist, M. Ortolano, Implementation of a graphene quantum Hall Kelvin bridge-on-a-chip for resistance calibrations. Metrologia **57**, 015007 (2020).

[3] A. R. Panna, I-F. Hu, M. Kruskopf, D. K. Patel, D. G. Jarrett, C.-I Liu, S. U. Payagala, D. Saha, A. F. Rigosi, D. B. Newell, C.-T. Liang, R. E. Elmquist, Graphene quantum Hall effect parallel resistance arrays. Phys. Rev. B **103**, 075408 (2021).

[4] J. Hu, A.F. Rigosi, M. Kruskopf, Y. Yang, B.-Y. Wu, J. Tian, A.R. Panna, H.-Y. Lee, S.U. Payagala, G.R. Jones, M.E. Kraft, D.G. Jarrett, K. Watanabe, T. Taniguchi, R.E. Elmquist, and D.B. Newell, Towards epitaxial graphene pn junctions as electrically programmable quantum resistance standards. Sci. Rep. **8**, 15018 (2018).

[5] S. Novikov, N. Lebedeva, J. Hamalainen, I. Iisakka, P. Immonen, A. J. Manninen, and A. Satrapinski, Mini array of quantum Hall devices based on epitaxial graphene. J. Appl. Phys. **119**, 174504 (2016).

[6] F. Delahaye, Series and parallel connection of multiterminal quantum Hall-effect devices. J. Appl. Phys. **73**, 7914 (1993).

[7] J. Hu, A.F. Rigosi, J.U. Lee, H.-Y. Lee, Y. Yang, C.-I. Liu, R.E. Elmquist, and D.B. Newell, Quantum transport in graphene p-n junctions with moiré superlattice modulation. Phys. Rev. B **98**, 045412 (2018).

[8] M. Woszczyna, M. Friedmann, T. Dziomba, T. Weimann, and F.J. Ahlers, Graphene pn junction arrays as quantum-Hall resistance standards. Appl. Phys. Lett. **99**, 022112 (2011).

[9] D. Patel, M. Marzano, C.-I Liu, H. M. Hill, M. Kruskopf, H. Jin, J. Hu, D. B. Newell, C.-T. Liang, R. Elmquist, A. F. Rigosi, Accessing ratios of quantized resistances in graphene p–n junction devices using multiple terminals. AIP Adv. **10**, 025112 (2020).

[10] Z. S. Momtaz, S. Heun, G. Biasiol, S. Roddaro, Cascaded Quantum Hall Bisection and Applications to Quantum Metrology. Phys. Rev. Appl. **14**, 024059 (2020).

[11] J. Park, W. S. Kim, Realization of 5h/e^2 with graphene quantum Hall resistance array. Appl. Phys. Lett. **116**, 093102 (2020).

[12] A. Lartsev, S. Lara-Avila, A. Danilov, S. Kubatkin, A. Tzalenchuk, and R. Yakimova, A prototype of RK/200 quantum Hall array resistance standard on epitaxial graphene. J. Appl. Phys. **118**, 044506 (2015).

[13] D. H. Chae, M. S. Kim, T. Oe, N. H. Kaneko, Series connection of quantum Hall resistance array and programmable Josephson voltage standard for current generation at one microampere. Metrologia **59**, 065011 (2022).

[14] S. M. Mhatre, N. T. M. Tran, H. M. Hill, C.-C. Yeh, D. Saha, D. B. Newell, A. R. Hight Walker, C.-T. Liang, R. E. Elmquist, A. F. Rigosi, Versatility of uniformly doped graphene quantum Hall arrays in series. AIP Adv. **12**, 085113 (2022).

[15] H. He, K. Cedergren, N. Shetty, S. Lara-Avila, S. Kubatkin, T. Bergsten, G. Eklund, Accurate graphene quantum Hall arrays for the new International System of Units. Nat. Commun. **13**, 6933 (2022).

[16] A. F. Rigosi, M. Marzano, A. Levy, H. M. Hill, D. K. Patel, M. Kruskopf, H. Jin, R. E. Elmquist, and D.B. Newell, Analytical determination of atypical quantized resistances in graphene pn junctions. Physica B: Condens. Matter **582**, 411971 (2020).

[17] C.-I Liu, D. S. Scaletta, D. K. Patel, M. Kruskopf, A. Levy, H. M. Hill, A. F. Rigosi, Analysing quantized resistance behaviour in graphene Corbino pn junction devices. J. Phys. D: Appl. Phys. **53**, 275301 (2020).

[18] A. F. Rigosi, A. L. Levy, M. R. Snure, N. R. Glavin, Turn of the decade: versatility of 2D hexagonal boron nitride. J. Phys. Mater. **4**, 032003 (2021).

[19] X. Wang, E. Khatami, F. Fei, J. Wyrick, P. Namboodiri, R. Kashid, A. F. Rigosi, G. Bryant, R. Silver, Experimental realization of an extended Fermi-Hubbard model using a 2D lattice of dopant-based quantum dots. Nat. Commun. **13**, 6824 (2022).

[20] H. M. Hill, A. F. Rigosi, S. Krylyuk, J. Tian, N. V. Nguyen, A. V. Davydov, D. B. Newell, A. R. Hight Walker, Comprehensive optical characterization of atomically thin NbSe2. Phys. Rev. B **98**, 165109 (2018).

[21] U. Wurstbauer, D. Majumder, S. S. Mandal, I. Dujovne, T. D. Rhone *et al*., Phys. Rev. Lett. **107**, 066804 (2011).

[22] D. S. Scaletta, S. M. Mhatre, N. T. M. Tran, C. H. Yang, H. M. Hill, Y. Yang, L. Meng, A. R. Panna, S. U. Payagala, R. E. Elmquist, D. G. Jarrett, D. B. Newell, and A. F. Rigosi, Optimization of graphene-based quantum Hall arrays for recursive star–mesh transformations. Appl. Phys. Lett. **123**, 153504 (2023).

[23] K. M. Yu, D. G. Jarrett, A. F. Rigosi, S. U. Payagala and M. E. Kraft, Comparison of Multiple Methods for Obtaining PΩ Resistances with Low Uncertainties. IEEE Trans. Instrum. Meas. **69**, 3729-3738 (2020).

[24] D. G. Jarrett, C.-C. Yeh, S. U. Payagala, A. R. Panna, Y. Yang, L. Meng, S. M. Mhatre, N. T. M. Tran, H. M. Hill, D. Saha, R. E. Elmquist, D. B. Newell, A. F. Rigosi, Graphene-Based Star-Mesh Resistance Networks. IEEE Trans. Instrum. Meas. **72**, 1502710 (2023).

[25] H. A. Sauer, Wye-Delta Transfer Standards for Calibration of Wide Range dc Resistance and dc Conductance Bridges. IEEE Trans. Instrum. Meas. **17**, 151-155 (1968).

[26] R. E. Scott, Linear circuits. 156-169, Addison-Wesley, (1960).

[27] A.E. Kennelly, Equivalence of triangles and stars in conducting networks. Electrical World and Engineer, **34**, 413–414 (1899).

[28] L. Versfeld, Remarks on Star-Mesh Transformation of Electrical Networks. Electron. Lett. **6**, 597-599 (1970).